\documentclass[aps,prl,twocolumn,superscriptaddress,
showpacs,nobalancelastpage,nobibnotes]{revtex4}
\usepackage[dvips]{epsfig}
\usepackage{graphicx}
\usepackage{amsmath}

\begin{document}

\title{GADZOOKS! Antineutrino Spectroscopy with Large
Water \v{C}erenkov Detectors}

\author{John F. Beacom}
\affiliation{NASA/Fermilab Astrophysics Center, Fermi National
Accelerator Laboratory, Batavia, Illinois 60510-0500}

\author{Mark R. Vagins}
\affiliation{Department of Physics and Astronomy, 4129 Reines Hall,
University of California, Irvine, CA 92697}

\date{25 September 2003}

\begin{abstract}
We propose modifying large water \v{C}erenkov detectors by the
addition of 0.2\% gadolinium trichloride, which is highly soluble,
newly inexpensive, and transparent in solution.  Since Gd has an
enormous cross section for radiative neutron capture, with $\sum
E_\gamma = 8$ MeV, this would make neutrons visible for the first time
in such detectors, allowing antineutrino tagging by the coincidence
detection reaction $\bar{\nu}_e + p \rightarrow e^+ + n$ (similarly
for $\bar{\nu}_\mu$).  Taking Super-Kamiokande as a working example,
dramatic consequences for reactor neutrino measurements, first
observation of the diffuse supernova neutrino background, Galactic
supernova detection, and other topics are discussed.
\end{abstract}

\pacs{95.55.Vj, 95.85.Ry, 14.60.Pq
\hspace{4.5cm} FERMILAB-Pub-03/249-A}


\maketitle


%
The Super-Kamiokande (SK) water \v{C}erenkov detector~\cite{SKNIM} has
very successfully observed solar~\cite{SKsolar},
atmospheric~\cite{SKatm}, and accelerator~\cite{SKk2k} neutrinos.
Even more subtle signals could be studied if antineutrinos were tagged
by the detection of neutrons following $\bar{\nu}_e + p \rightarrow
e^+ + n$ (similarly for $\bar{\nu}_\mu$), as the coincidence detection
greatly reduces backgrounds from radioactivities and other neutrino
reactions.  We propose that this could be accomplished by the addition
of $0.2\%$ gadolinium trichloride (GdCl$_3$) to the water, and refer
to the resulting detector as GADZOOKS! ({\it Gadolinium Antineutrino
Detector Zealously Outperforming Old Kamiokande, Super!}).  While this
ability exists in $\lesssim 1$ kton detectors, it has never been
achieved on the scale of the 22.5 kton fiducial mass of SK, nor in a
light water detector.  The proposed use of GdCl$_3$ dissolved in water
is also novel.  We discuss SK as a concrete example to show
feasibility, but the proposed technique has much more general
applicability.


{\bf Neutron Capture on Gadolinium.---}
Neutrons in water quickly lose energy by collisions with free protons
(and oxygen nuclei); once thermal energies are reached, the neutron
continues to undergo collisions, changing its direction, but on
average not its energy, until it is captured~\cite{invbeta}.  The
cross section for thermal neutron capture on natural Gd is 49000
barns, compared to 0.3 barns on free protons~\cite{isotopes}.  With
the proposed 0.2\% admixture by mass of GdCl$_3$ in water, 90\% of
neutron captures are on Gd (8 MeV gamma cascade), 0.2\% on Cl (8.6 MeV
gamma cascade), and the rest on H (2.2 MeV gamma, not detectable in
SK).  After thermalization, capture occurs in about 20 $\mu$s (about
10 times faster than in pure water) and about 4 cm; both are slightly
increased by the pre-thermalization phase.  Compared to typical time
and distance separations between events in SK, as well as the position
resolution, these are exceedingly small.

Since SK was first proposed, the price of the water soluble salt
GdCl$_3$ has fallen three orders of magnitude, most of that in the
past few years.  The current price of 99.99\% pure GdCl$_3 \cdot
X$H$_2$O from the Stanford Materials Corporation is about \$3/kg.  The
properties of GdCl$_3$ are discussed in more detail below.  Neutron
capture on Gd leads to 8 MeV shared among $3-4$ gammas.  In
scintillators, where Gd is used routinely, the summed gamma energy is
relevant.  However, in water what matters are the electrons
Compton-scattered above the \v{C}erenkov threshold by relatively hard
gammas.  The detectable light following neutron capture on Gd (in thin
foils) in possible discrete counters was carefully
simulated~\cite{SNOGd} for the SNO heavy water \v{C}erenkov detector.
The equivalent single electron energy was found to peak at about 5
MeV, and range over $3-8$ MeV.  This spread reflects the intrinsic
variation in the gamma cascades and the detector energy resolution
(the simulation had 5 photoelectrons per MeV of electron energy,
compared to 6 in SK-I).

At energies as low as 3 MeV, background singles rates are much too
high to detect neutron captures in isolation, as required in SNO.
However, using the detection reaction $\bar{\nu}_e + p \rightarrow e^+
+ n$ and requiring a very tight time and position coincidence between
the positron and the following neutron will allow detection of nearly
all captures on Gd.  Due to continuing improvements in online data
acquisition and filtering, it is expected that SK-III, which will
begin operations in mid-2006, will trigger at 100\% efficiency at 3
MeV and above, with good trigger efficiency down to 2.5 MeV.  The rate
of accidental coincidences is vanishing, since the number of candidate
solar neutrino events is $\sim 1$/yr/ton in SK.  Gamma cascades also
produce more isotropic light than other backgrounds, which aids in
their identification.


{\bf Neutron Rates in SK.---}
The ambient neutron rate in the SK fiducial volume is unknown, as
neutrons and their captures on free protons are invisible.  Though a
high background neutron rate could be tolerated for the coincidence
signal $\bar{\nu}_e + p \rightarrow e^+ + n$, we do not want to
compromise the $5-20$ MeV solar neutrino singles signal by making
neutron captures visible.  There are $\sim 10$ signal events per day
along the solar direction, and $\sim 100$ background events per day
over all directions.  After examining a wide variety of potential
sources of ambient neutrons, we find that they are not a problem.

Cosmic ray muon spallation in SK produces a very steeply falling
spectrum of $\sim 10^5$ neutrons/day; their production is of
interest~\cite{spalln}.  Spallation also produces beta-unstable
daughters that form a significant background for solar neutrinos;
after cuts following muons, $\lesssim 100$ spallation betas/day
survive.  Since these cuts veto events within seconds and meters of
muon tracks, they will be much more effective on neutrons, which
capture in much less time and distance.  Neutrons from spallation
events in the surrounding rock are stopped by 4.5 m of water shielding
surrounding the fiducial volume.  Our conclusions are supported by
data from KamLAND, which is adjacent to SK and has similar
shielding~\cite{KamLAND}.  Dissolved Gd would also be present in the
optically isolated outer detector, and the neutron capture rate there
could be as high as $10^3$ Hz, due to neutrons from the surrounding
rock~\cite{SKNIM}.  However, captures on Gd would not produce enough
light to trigger the sparse outer detector.

The rate of $\bar{\nu}_e + p \rightarrow e^+ + n$ interactions in SK
from nuclear reactors is $\simeq 30$/day.  For the majority of events
the positron would be detectable in coincidence with the neutron (even
without that, the excess singles rate due to neutron captures could be
detected).  For the lowest energy reactor antineutrinos, as well as
those originating from U/Th decays in Earth, the positron energy is
too low to trigger SK, and the neutron would appear in isolation.  The
rate due to U/Th decays is expected to be $\sim 4$/day.  Atmospheric
neutrino neutral-current events may be otherwise invisible if only a
neutron is scattered, a relatively common event, at a rate of $\sim
2$/day.

At the present U/Th/Rn concentrations in SK, $\lesssim 1$ neutron/day
in total is produced by the following processes: spontaneous fission
of $^{238}$U; $(\alpha,n)$ reactions on $^{2}$H, $^{17}$O, and
$^{18}$O; and $^{2}$H$(\gamma,n)$; estimated by scaling from SNO
results~\cite{SNONC}.  The GdCl$_3$ additive must meet radiopurity
standards about $10^3$ times less stringent than for the SK water.
Initial test samples, for which no special care was taken, were
measured by mass spectroscopy to have $[^{238}{\rm U}] \simeq 10^{-8}$
g/g and $[^{232}{\rm Th}] \simeq 10^{-10}$ g/g~\cite{NandM}.  Assuming
secular equilibrium in the decay chains, the beta and neutron rates in
SK would remain similar to present values if the samples were purified by
a factor of 100.  The Palo Verde experiment obtained Gd 10 times more
radiopure than our initial samples, and the SK water system reduces
U/Th/Rn by orders of magnitude from the original mine water.  We are
confident that the desired radiopurity is easily obtainable (as was
Ref.~\cite{SNOGd}).

Natural Gd contains 0.2\% $^{152}$Gd, which alpha decays ($T_{1/2} =
10^{14}$ yr, $T_\alpha = 2.1$ MeV)~\cite{isotopes}.  With 100 tons of
GdCl$_3$ in SK, the decay rate is $\sim 10^{10}$/day.  These alphas
are invisible in SK, but their introduction may initiate $^{17}{\rm
O}(\alpha,n)$ and $^{18}{\rm O}(\alpha,n)$ reactions.  Using the alpha
stopping power and the measured cross sections~\cite{alphaOxy}, the
neutron production rate is $\sim 1$/day.  Lanthanide contaminants
($\lesssim 10^{-4}$) and their decays can also be ignored.


{\bf Reactor Neutrinos.---}
The total $\bar{\nu}_e + p \rightarrow e^+ + n$ rate in SK from
reactors can be scaled from the KamLAND rate~\cite{KamLAND}, and is
$\simeq 30$/day after oscillations.  The threshold for solar neutrinos
in SK-I was $E_e = 5$ MeV, and SK-III should be even better due to a
lowered trigger threshold and much-improved offline reconstruction
algorithms currently being evaluated in SK-II.  This was the {\it
analysis} threshold for single events.  The trigger is efficient to
much lower energies (as low as 3.5 MeV in SK-I), where radioactivity
backgrounds overwhelm the solar singles rate.  However, for the
coincidence signal, it should be possible to identify real events as
low as $E_e = 2.5$ MeV, covering most of the reactor spectrum.  The
present KamLAND analysis threshold is $E_{vis} = T_e + 2 m_e = 2.6$
MeV, corresponding to $E_e = 2.1$ MeV in SK.  Measurement of the
positron energy determines the neutrino energy, since $E_\nu \simeq
E_e + 1.3$ MeV, and additionally there is a weak directional
correlation~\cite{invbeta}.

GADZOOKS! would have the advantage of much larger statistics, with
about 50 times more fiducial mass.  The very high rate would allow the
flux to be monitored on yearly basis with about 1\% statistical error,
likely allowing new tests of neutrino oscillation parameters as
reactors at different distances go through on/off cycles.  The
expected reactor spectrum is shown in Fig.~1.  The energy resolution
in SK is about 6 times worse than in KamLAND, so that spectral
distortions (not shown) will be smeared, though high statistics may
still reveal them.  Resolution is why the positron spectrum extends to
$E_e = 12$ MeV, even though the neutrino spectrum~\cite{reactor}
plummets beyond $E_\nu = 8$ MeV.

In KamLAND, fast neutrons from the walls can scatter protons
(producing scintillation light) and then capture, mimicking the
coincidence signal; in SK, the struck protons are invisible.  Muon
spallation produces three isotopes ($^{8}$He, $^{9}$Li, and $^{11}$Li)
that beta decay to excited daughter states that decay by neutron
emission~\cite{isotopes}.  In KamLAND, spallation cuts control these
to be $\lesssim 2\%$ of the reactor signal.  Spontaneous fission of
$^{238}$U and atmospheric neutrino neutral-current interactions can
both produce multiple neutron events that can mimic the reactor
signal, but the rates are small.


{\bf Diffuse Supernova Neutrino Background.---}
From a supernova in the nearest galaxy, Andromeda, SK would detect
$\sim 1$ event.  While more distant galaxies have lower fluxes, there
are many more of them, and the flux from all previous core-collapse
supernovae in the universe may be detectable.  We refer to this as the
{\it Diffuse Supernova Neutrino Background} (DSNB), and not the more
common ``relic supernova neutrinos,'' since the latter causes
confusion with ``relic'' big bang neutrinos.  All neutrino flavors are
produced, but $\bar{\nu}_e$ is the easiest to detect, via $\bar{\nu}_e
+ p \rightarrow e^+ + n$ on free protons.  The strongest limits are
from SK, based on the electron spectrum above 18 MeV (at 18 MeV,
the measured rate is $\sim 1$/22.5 kton/yr/MeV)~\cite{Malek}.
The DSNB spectrum is a convolution of a single supernova spectrum with
the supernova rate as a function of $z$, with neutrino energies
redshifted as $E_\nu/(1 + z)$.  The Kaplinghat, Steigman, and Walker
(KSW) model pushed uncertainties in the direction of producing the
largest reasonable DSNB flux~\cite{Kaplinghat}.  In the relevant
energy range, $E = 10-30$ MeV, other models have nearly the same shape
but differ mostly in normalization; also, several uncertainties are
minimized.  The supernova rate is reasonably known in the relevant
range $z \lesssim 1$ (where it rises by $\sim 10$ over the $z = 0$
rate)~\cite{Kaplinghat,modern}.  Uncertainties on cosmological and
neutrino oscillation parameters no longer play a significant
role~\cite{modern}, the latter especially if realistic neutrino
temperatures~\cite{Keil} are used.  The DSNB detection cross
section~\cite{invbeta} may be treated at lowest order at present.

In Fig.~1, we show a range of DSNB spectra.  The upper edge of the
band is set by the SK limit~\cite{Malek} (0.6 of KSW), and the lower
edge by modern models~\cite{modern} (0.2 of KSW).  The
background-limited SK search~\cite{Malek} will gain the required
factor of 3 in sensitivity in about 40 years.  In GADZOOKS!, this
sensitivity would be available immediately.  Requiring neutron
detection would dramatically lower the backgrounds below 18 MeV, where
the spallation beta singles rate rises rapidly.  As the threshold is
lowered, the atmospheric neutrino backgrounds fall and the DSNB signal
rises.  We calculate that the atmospheric~\cite{atmlow} backgrounds
can be reduced by $\sim 5$ from the measured rates~\cite{Malek} by
rejecting events with a preceding nuclear gamma~\cite{Kolbe} or
without a following neutron.  Further rejection (not shown) is likely
possible by requiring a small position separation between prompt and
delayed events, since DSNB events produce much less energetic
neutrons.  The number of DSNB events expected in GADZOOKS!  is about
$2-6$ per year above 10 MeV.  Uncertainties smaller than the
corresponding Poisson uncertainty can be ignored.

Detection of the DSNB would be an extremely important scientific
milestone.  It could be the first detection of neutrinos from
significant redshifts $z \lesssim 1$, and the second detection of
supernova neutrinos.  With the exception of SN 1987A in the Large
Magellanic Cloud, a close companion of our Galaxy, neutrinos have
never been detected from farther than the Sun.  The DSNB flux is
proportional to the rate of all core-collapse supernovae, including
optically dark ``failed'' supernovae that collapse to black holes.
The DSNB spectrum shape would also provide a crucial calibration for
numerical supernova models, since in relatively few years, the sample
of neutrinos from SN 1987A could be surpassed

\begin{figure}
\includegraphics[width=3.25in]{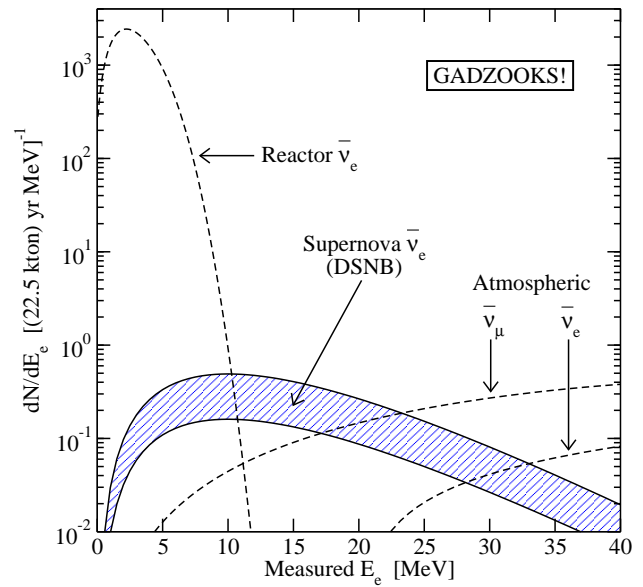}
\caption{Spectra of low-energy $\bar{\nu}_e + p \rightarrow e^+ + n$
coincidence events and the sub-\v{C}erenkov muon background.
We assume full efficiencies, and include energy resolution and neutrino
oscillations.  Singles rates (not shown) are efficiently suppressed.}
\end{figure}


{\bf Galactic Supernova.---}
Supernovae in our Galaxy are expected about 3 times per century, and
SK would observe $\sim 10^4$ events at a typical distance of 10 kpc.
The ability to cleanly identify the dominant $\bar{\nu}_e + p
\rightarrow e^+ + n$ events would be extremely important for studying
the remaining reactions, notably $\nu_e + ^{16}{\rm O} \rightarrow e^-
+ ^{16}{\rm F}$, which is exquisitely sensitive to the $\nu_e$
temperature and hence neutrino mixing~\cite{oxygenCC}.  Hundreds of
$\nu_e$ events could be observed, far more than in any other detector.
The neutral-current events on $^{16}$O that lead to gamma and/or
neutron emission would be much better identified, of key importance
for measuring the $\nu_\mu/\nu_\tau$ temperature~\cite{oxygenNC}.
Even in the forward angular cone, inverse beta events dominate
neutrino-electron scattering events~\cite{SNpointing}.  Isolating
those events would be very useful: it would help detect the
neutronization burst $\nu_e$ events; it would improve the pointing to
the supernova by a factor of about 2, down to about
$2^\circ$~\cite{SNpointing}; and it would allow better spectral
studies.  Using timing information alone, SK could immediately
recognize a supernova as genuine by the unique time structure of the
events: almost all events in pairs separated by tens of microseconds,
much shorter than the separation between subsequent neutrino
interactions.  Neutron detection would improve the ability to study
bursts out to very late times, or to detect faint bursts.


{\bf Other Physics.---}
The solar $\bar{\nu}_e$ flux is $\lesssim 1\%$ of the predicted
$\nu_e$ flux~\cite{Gando}.  Requiring a neutron coincidence would
greatly reduce backgrounds, and the sensitivity would be better than
about 0.01\%; the weak directional information~\cite{invbeta} may
allow even better sensitivity.  For atmospheric and accelerator
neutrinos, the ability to detect neutron captures delayed from the
neutrino interaction would shed new light on the hadronic final state.
This would be useful for separating (especially sub-GeV) neutrinos and
antineutrinos (which preferentially eject protons and neutrons,
respectively) and for probing the type of neutrino interaction.  For
accelerator neutrinos, neutrino-neutron elastic scattering will be a
significant new neutral-current channel.  For proton decay, neutron
detection may reduce atmospheric neutrino backgrounds, as well as aid
the study of bound-nucleon decay modes (including invisible modes) in
which the final nucleus decays with an emitted neutron.


{\bf R\&D on GADZOOKS!---}
Many of the properties of GdCl$_3$ are already well known.  It is
naturally highly water soluble, in concentrations up to 50\%.  In
contrast, an elaborate chemistry is needed to suspend even modest
amounts of Gd in liquid scintillator.  GdCl$_3$ is stable,
non-reactive, and easy to handle in bulk quantities.  Years of human
and animal studies have demonstrated its non-toxicity~\cite{nontoxic},
and it is commonly employed as a safe and effective contrasting agent
in human MRI subjects.  Still, because of the novel use we are
proposing for GdCl$_3$, some new, specialized information is desired.

In our calculations, we treated neutron capture on Gd, but otherwise
ignored the effects of GdCl$_3$ on the detector (note that Haxton
discussed adding $5-10\%$ admixtures of various other salts to SK as
solar neutrino targets, but not for neutron capture~\cite{Haxton}).
Three key R\&D issues are currently being addressed~\cite{ADRP}.
First, while preliminary measurements indicate that the light
absorption length over the \v{C}erenkov frequency range in Gd-loaded
SK water remains $\gtrsim 100$ m~\cite{NandM}, its complete light
attenuation characteristics (including scattering) over these long
distances will be evaluated.  Second, the physical effects of GdCl$_3$
on detector components will be investigated through extended exposure
of samples of these materials, as well as through the use of
established accelerated aging techniques.  Finally, an operational
replica of the SK water filtration system will be constructed in order
to determine the most effective method of filtering the water of other
impurities without incurring unacceptable losses or concentration
variations of the 0.2\% GdCl$_3$ solute in the process.


{\bf Conclusions.---}
We propose GADZOOKS!, a large water \v{C}erenkov detector that would
allow neutron detection by radiative capture on a dissolved gadolinium
salt.  Extensive research and preliminary R\&D, briefly reported here,
strongly support the feasibility of this technique, and a program of
focused R\&D has just been funded~\cite{ADRP}.  The new ability to
detect neutrons in a large light water detector would allow the clear
identification of $\bar{\nu}_e$ by the coincidence detection reaction
$\bar{\nu}_e + p \rightarrow e^+ + n$ (similarly for $\bar{\nu}_\mu$).
An entirely new program of antineutrino spectroscopy would be opened,
with important implications for reactor, solar, supernova,
atmospheric, and accelerator neutrinos.  The prospects for the first
detection of the diffuse supernova neutrino background are
particularly exciting.  This could be the first detection of neutrinos
from cosmological distances (and certainly the largest $L/E$), and the
second detection of supernova neutrinos.  In just a few years, the
yield from SN 1987A could be surpassed.  Unlike previous neutron
detection techniques, ours is scalable at reasonable expense.
Megaton-scale water detectors with GdCl$_3$ could observe hundreds of
DSNB events per year, allowing stringent tests of the black hole
formation rate and supernova neutrino spectra.  Other physics topics
would enjoy similar benefits.


{\bf Acknowledgments.---}
We thank Y. Suzuki, M. Nakahata, H. Sobel, M. Koshiba, and T. Kajita
for strong support and sage advice; P. Vogel and S. Zeller for key
assistance; and S. Brice, T. Haines, A. Piepke, H. Robertson, and
R. Svoboda for helpful discussions.  JFB was supported by DOE
DE-AC02-76CH03000 and NASA NAG5-10842, and MRV by DOE
DE-FG03-91ER40679 and DE-FG02-03ER41266.




\begin{thebibliography}{99}


\bibitem{SKNIM}
Y.~Fukuda {\it et al.}, 
Nucl.\ Instrum.\ Meth.\ A {\bf 501}, 418 (2003).

\bibitem{SKsolar}
S.~Fukuda {\it et al.}, 
Phys.\ Lett.\ B {\bf 539}, 179 (2002).

\bibitem{SKatm}
Y.~Fukuda {\it et al.}, 
Phys.\ Rev.\ Lett.\  {\bf 81}, 1562 (1998).

\bibitem{SKk2k}
M.~H.~Ahn {\it et al.}, 
Phys.\ Rev.\ Lett.\  {\bf 90}, 041801 (2003).

\bibitem{invbeta}
P.~Vogel and J.~F.~Beacom,
Phys.\ Rev.\ D {\bf 60}, 053003 (1999).

\bibitem{isotopes}
R. B. Firestone and V. S. Shirley, {\it Table of Isotopes}
(John Wiley, New York, 1996).

\bibitem{SNOGd}
C.~K.~Hargrove, I.~Blevis, D.~Paterson and E.~D.~Earle,
Nucl.\ Instrum.\ Meth.\ A {\bf 357}, 157 (1995).

\bibitem{spalln}
F.~Boehm {\it et al.}, 
Phys.\ Rev.\ D {\bf 62}, 092005 (2000);
V.~A.~Kudryavtsev, N.~J.~Spooner and J.~E.~McMillan,
Nucl.\ Instrum.\ Meth.\ A {\bf 505}, 683 (2003).

\bibitem{KamLAND}
K.~Eguchi {\it et al.}, 
Phys.\ Rev.\ Lett.\  {\bf 90}, 021802 (2003).

\bibitem{SNONC}
Q.~R.~Ahmad {\it et al.}, 
Phys.\ Rev.\ Lett.\  {\bf 89}, 011301 (2002);
Phys.\ Rev.\ Lett.\  {\bf 89}, 011302 (2002).

\bibitem{NandM}
M. Nakahata and C. Mitsuda, private communication.

\bibitem{alphaOxy}
J.K. Bair and F.X. Haas,
Phys. Rev. C {\bf 7}, 1356 (1973).

\bibitem{reactor}
V.~I.~Kopeikin, L.~A.~Mikaelyan and V.~V.~Sinev,
Phys.\ Atom.\ Nucl.\  {\bf 60}, 172 (1997)
[Yad.\ Fiz.\  {\bf 60}, 230 (1997)].

\bibitem{Malek}
M.~Malek {\it et al.}, 
Phys.\ Rev.\ Lett.\  {\bf 90}, 061101 (2003).

\bibitem{Kaplinghat}
M.~Kaplinghat, G.~Steigman and T.~P.~Walker,
Phys.\ Rev.\ D {\bf 62}, 043001 (2000).

\bibitem{modern}
S.~Ando, K.~Sato and T.~Totani,
Astropart.\ Phys.\  {\bf 18}, 307 (2003);
M.~Fukugita and M.~Kawasaki,
Mon.\ Not.\ Roy.\ Astron.\ Soc.\  {\bf 340}, L7 (2003).

\bibitem{Keil}
M.~T.~Keil, G.~G.~Raffelt and H.~T.~Janka,
Astrophys.\ J.\  {\bf 590}, 971 (2003).

\bibitem{atmlow}
G.~Barr, T.~K.~Gaisser and T.~Stanev,
Phys.\ Rev.\ D {\bf 39}, 3532 (1989).

\bibitem{Kolbe}
E.~Kolbe, K.~Langanke and P.~Vogel,
Phys.\ Rev.\ D {\bf 66}, 013007 (2002).

\bibitem{oxygenCC}
W.~C.~Haxton,
Phys.\ Rev.\ D {\bf 36}, 2283 (1987).

\bibitem{oxygenNC}
K.~Langanke, P.~Vogel and E.~Kolbe,
Phys.\ Rev.\ Lett.\  {\bf 76}, 2629 (1996); 
J.~F.~Beacom and P.~Vogel,
Phys.\ Rev.\ D {\bf 58}, 053010 (1998);
Phys.\ Rev.\ D {\bf 58}, 093012 (1998).

\bibitem{SNpointing}
J.~F.~Beacom and P.~Vogel,
Phys.\ Rev.\ D {\bf 60}, 033007 (1999);
R.~Tomas {\it et al.},
hep-ph/0307050.

\bibitem{Gando}
Y.~Gando {\it et al.}, 
Phys.\ Rev.\ Lett.\  {\bf 90}, 171302 (2003).

\bibitem{nontoxic}
T.J. Haley {\it et al.},
Brit. J. Pharmacol. {\bf 17}, 526 (1961);
S. Yoneda {\it et al.},
Fundam. Appl. Toxicol. {\bf 28}, 65 (1995).

\bibitem{Haxton}
W.~C.~Haxton,
Phys.\ Rev.\ Lett.\  {\bf 76}, 1562 (1996).

\bibitem{ADRP}
M.R. Vagins (Principal Investigator), Department of Energy 2003
Advanced Detector Research Program grant.

\end{thebibliography}
\end{document}